# Function Estimation Using Data-Adaptive Kernel Smoothers—How Much Smoothing?

K. S. Riedel, A. Sidorenko, and James R. Matey

---

---

# FUNCTION ESTIMATION USING DATA-ADAPTIVE KERNEL SMOOTHERS—HOW MUCH SMOOTHING?

**K. S. Riedel and A. Sidorenko**

We consider a common problem in physics: how to estimate a smooth function given noisy measurements. We assume that the unknown signal is measured at $N$ different times $\{t_i: i = 1, \ldots N\}$ and that the measurements $\{y_i\}$ have been contaminated by additive noise. Thus, the measurements satisfy $y_i = g(t_i) + \varepsilon_i$, where $g(t)$ is the unknown signal and $\varepsilon_i$ represents random errors. For simplicity, we assume that the errors are independent and have zero mean and uniform variance $\sigma^2$.

As an example, we consider a chirp signal: $g(t) = \sin(4\pi t^2)$. This signal is called a chirp because its "frequency" is growing linearly: $(d/dt)\{\text{phase}\} = 8\pi t$, which corresponds to the changing pitch in a bird's chirp. Figure 1 plots the chirp over two periods. Superimposed on the chirp is a point random realization of the noisy signal with $\sigma = 0.5$.

A simple estimator of the unknown signal is a local average:

$$\hat{g}(t_i) = \frac{1}{2L+1} \sum_{j=-L}^{L} y_{i+j} \ , \qquad (1)$$

where we assume that the sampling times are uniformly spaced. We denote the sampling rate, $t_{i+1} - t_i$, by $\Delta$, and define the normalized kernel halfwidth $h \equiv (L\Delta)$. The ^ notation denotes the estimate of the unknown function.

When $L = 0$, the estimate is simply the point value: $\hat{g}(t_i) = y_i$, which has variance $\sigma^2$. By averaging over $2L + 1$ independent measurements, the variance of the estimate is reduced by a factor of $1/(2L + 1)$ to $\sigma^2/(2L + 1)$.

As we increase the averaging halfwidth $L$, the variance of the estimate will decrease. However, the local average in Eq. (1) includes other data points that systematically differ from $g(t_i)$. As a result, the local average has a systematic bias error in estimating $\hat{g}(t_i)$:

$$\mathbf{E}[\hat{g}(t_i)] = \frac{1}{2L+1} \sum_{j=-L}^{L} g(t_{i+j}) \neq g(t_i) \ ,$$

where $\mathbf{E}$ is the expectation of the estimate. As $L$ increases, the variance decreases while the systematic error normally increases. This is a typical example of "bias-versus-variance trade-off" in data analysis.

Figure 2 plots the local average estimate of Eq. (1) for three different values of the halfwidth $L$: 4, 9, and 14. The averaged curves change discontinuously as measurements are added and deleted from the average. As a result, the averaged curve is harsh and unappealing. The simple average using a rectangular weighting has other disadvantages, such as mak-

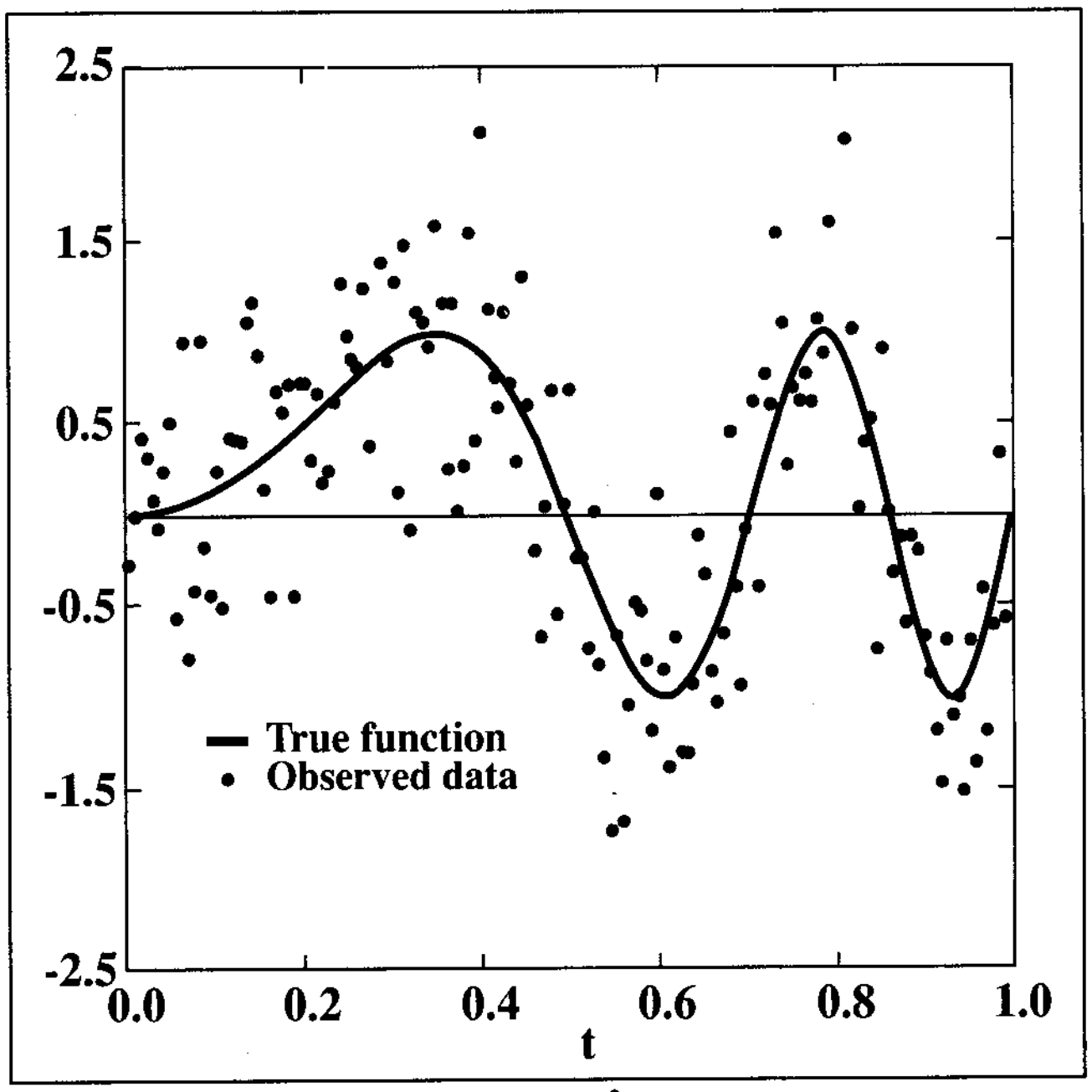

*Figure 1. Chirp signal: $g(t) = \sin(4\pi t^2)$. A random realization of 150 data points is superimposed with $\sigma = 0.5$.*

*K.S. Riedel is at the Courant Institute, New York University, 251 Mercer St., New York, NY 10012-1185. E-mail: riedel@cims.nyu.edu*

*A. Sidorenko is at the Courant Institute. E-mail: sidorenk@cims.nyu.edu*

ing less-accurate estimates. Thus, we consider more-general kernel estimates by allowing an arbitrary weighting of the measurements:

$$\hat{g}(t_i) = \sum_{j=-L}^{L} w_j(t_i)\, y_{i+j}\,, \qquad (2)$$

where the $w_j$ are weights. Typically, the weights are given by a scale function, $\kappa(t)$: $w_j = c\kappa[(t_i - t_j)/h]$. The constant $c$ is chosen such that $\sum_j w_j = 1$. We use a parabolic weighting, $\kappa(t) = (3/4)(1 - t^2)$, for $|t| \le h$ and zero otherwise. Near the ends of the data, we modify the kernel [see Eq. (A3)]. The normalized halfwidth $h$ is still a free parameter that determines the strength of the smoothing, and $h$ can depend on the estimation point $t_i$.

Figure 3 plots the local average estimate using the parabolic weighting for three different values of the halfwidth $h$: 0.04, 0.08, and 0.12, corresponding to $L$: 5, 11, and 17. The smoothed curves are continuous and are more aesthetic than those of Fig. 2. The $h = 0.04$ average has a lot of random jitter, indicating that the variance of the estimate is still large. The height of the second maximum and the depth of the two minima have been appreciably reduced in the $h = 0.08$ and the $h = 0.12$ averages. In fact, the $h = 0.12$ average misses the second minimum at $t = 0.935$ entirely. Because the estimated curve significantly depends on $h$, Fig. 3 shows the need for care in choosing the smoothing parameter.

One of the main issues discussed in this tutorial is how to pick $h$. In practice, to minimize artificial wiggles and nonphysical aspects, most scientists adjust the smoothing parameter according to their physical intuition. Using intuition makes the analysis subjective and often leads to suboptimal fits.

We would like to choose $h$ in such a way as to minimize the fitting error. Unfortunately, the fitting error depends on the unknown signal and therefore is unknown. We consider multiple stage estimators that automatically determine the smoothing parameter. In the first stage, we estimate the fitting error and then choose the smoothing parameter to minimize this empirical estimate of the error. The combined estimate is nonlinear in the measurements and automatically adapts the local smoothing to the curvature of the unknown function.

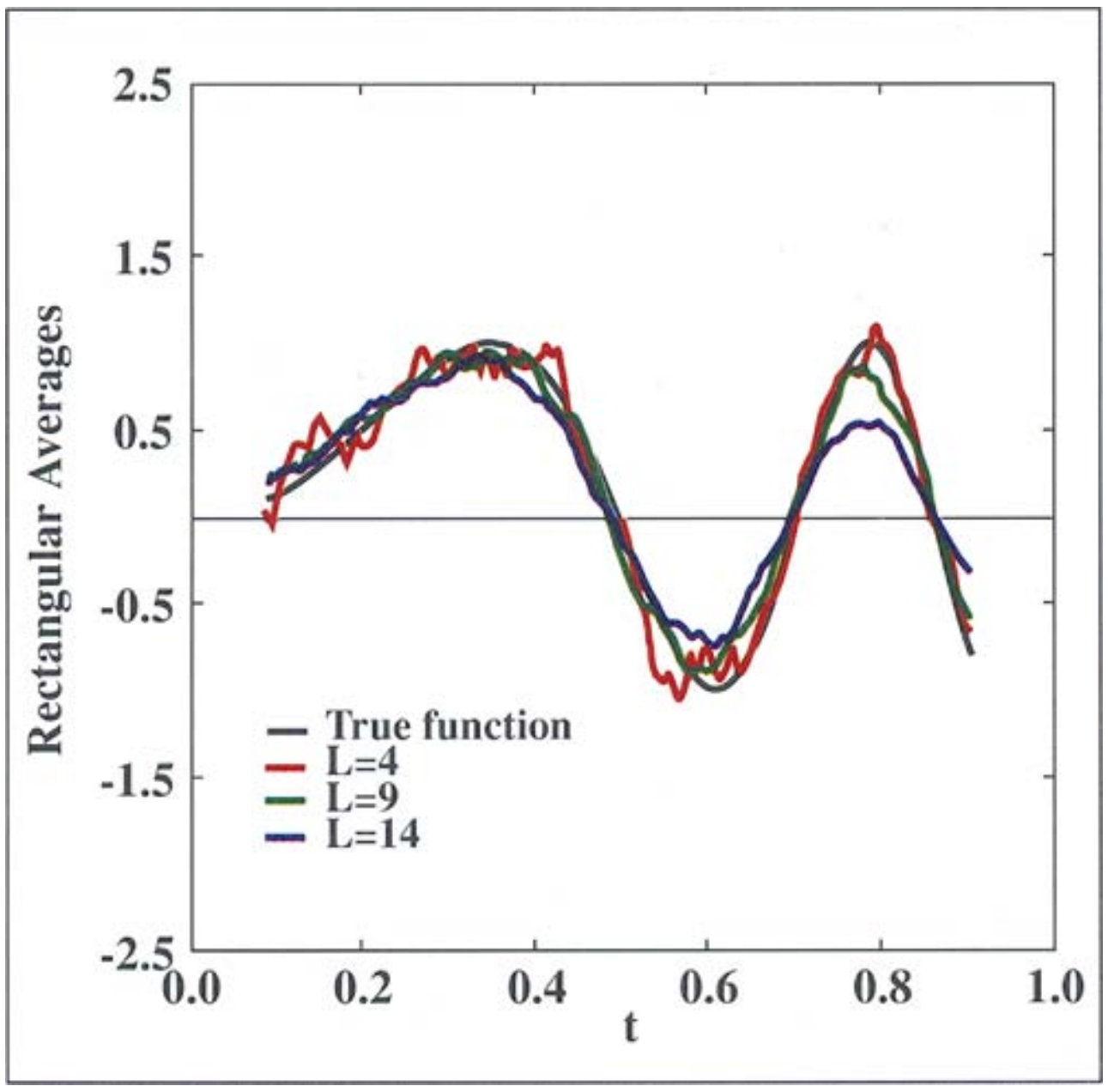

*Figure 2. Three kernel-smoother estimates of the randomized chirp signal using a rectangular window. The curves are actually discontinuous at the points where measurements are added and deleted.*

## Bias-versus-variance trade-off

We now give a local-error analysis of kernel smoothers, based on a Taylor-series approximation of the unknown function. Essentially, the sampling rate is required to be rapid in comparison with the time scale on which the unknown function is varying.

The advantage of the weighted local average is that the variance of the estimate is reduced. To see this variance reduction, we rewrite Eq. (2) as

$$\hat{g}(t_i) = \sum_{j=-L}^{L} w_j\,[g(t_{i+j}) + \varepsilon_{i+j}]\,. \qquad (3)$$

Because the errors are independent, the resulting variance is

$$\mathbf{Var}[\hat{g}(t_i)] = \sum_{j=-L}^{L} w_j^2\,\mathbf{Var}[\varepsilon_{i+j}] = \sigma^2 \sum_{j=-L}^{L} w_j^2\,. \qquad (4)$$

For the uniform weighting of Eq. (1), $w_j = 1/(2L+1)$ and the variance of the estimate is $\sigma^2/(2L+1)$. This same scaling of the variance holds for the more general class of scale function kernels, $\kappa(t)$: $w_j = c\kappa\,[(t_i - t_j)/h]$. The disadvantage is that averaging causes systematic error. We define the bias error to be

$$\text{Bias}[\hat{g}(t_i)] = \mathbf{E}\,[\hat{g}(t_i)] - g(t_i) = \sum_{j=-L}^{L} w_j\, g(t_{i+j}) - g(t_i)\,, \quad (5)$$

where **E** denotes the expectation. As the averaging halfwidth $L$ increases, $|g(t_{i+L}) - g(t_i)|$ will normally increase, so the bias error will generally grow with increasing amounts of averaging. The expected square error (ESE) is

$$\text{ESE}(\hat{g}) = [\text{Bias}]^2 + \text{Variance} \quad (6)$$

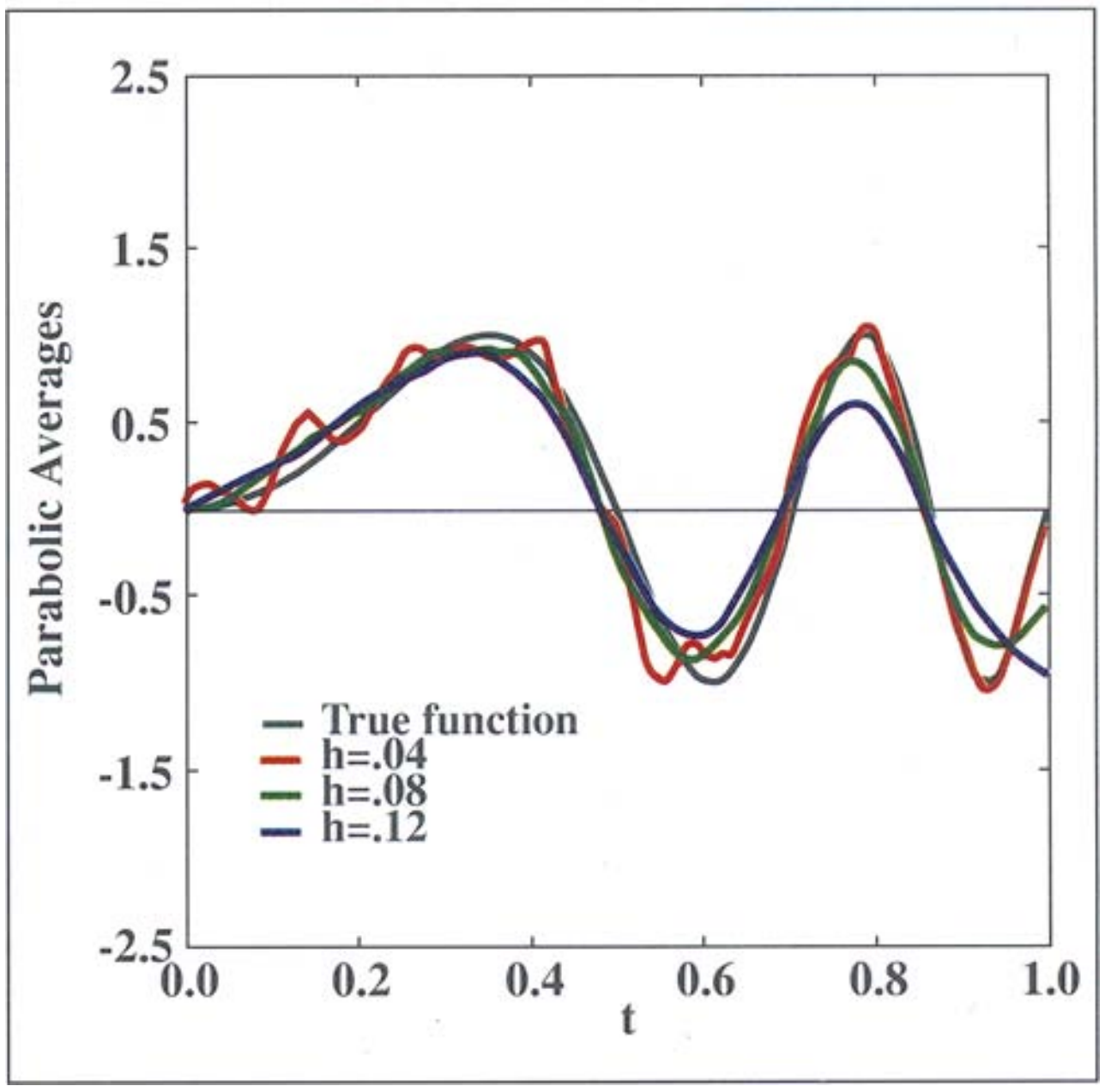

*Figure 3. Smoothed estimates of the randomized chirp signal using a parabolic kernel with three different values of the halfwidth h: 0.04, 0.08, 0.12, corresponding to L: 5, 11, 17. The parabolic weighting makes the curves continuous and more aesthetically pleasing than those of Fig. 2. The h = 0.04 average still has a lot of random jitter, indicating that the variance of the estimate is still large. The heights of the last three extrema have been reduced in the h = 0.08 and the h = 0.12 averages. The h = 0.12 average misses the last minimum entirely.*

Figure 4 plots the ESEs of the three parabolic kernel averages. The smallest halfwidth has the largest ESE for $t \le 0.5$. As time progresses, $g(t)$ oscillates more frequently. As a result, the bias error and the corresponding ESE for all three estimates oscillate more rapidly and increase. For $0.75 \le t \le 1.0$, the smallest halfwidth is the most reasonable, illustrating that the halfwidth of the kernel smoother should decrease when the unknown function varies more rapidly.

Our goal in this tutorial is to minimize the ESE of the kernel estimate of $\hat{g}(t)$. Since the ESE is unknown, we estimate the ESE and then optimize the kernel halfwidth with respect to the estimated ESE.

## Local error and optimal kernels

To understand the systematic error from smoothings, we make a Taylor series expansion of $g(t)$ about $t_i$: $g(t) = g(t_i) + g'(t_i)(t - t_i) + g''(t_i)[(t - t_i)^2/2]\ldots$. We make this Taylor-series expansion over the kernel halfwidth: $[t_{i-L}, t_{i+L}]$. For this expansion to be valid, the signal $g(t)$ must evolve slowly with respect to this averaging time, $t_{i+L} - t_{i-L}$.

We assume that the kernel weights satisfy the moment conditions:

$$\sum_{j=-L}^{L} w_j = 1\,, \quad \sum_{j=-L}^{L} (t_{i+j} - t_i) w_j = 0\,. \quad (7)$$

We define the following moments :

$$B_L(t_i) = \frac{1}{2h^2} \sum_{j=-L}^{L} (t_{i+j} - t_i)^2\, w_j\,, \quad C_L = L \sum_{j=-L}^{L} w_j^2\,. \quad (8a)$$

We now consider the sampling limit where the number of measurements $N$ in a fixed time interval tends to infinity and the sampling time $\Delta$ tends to zero. In this fast-sampling limit, when the weights are given by a scale function, $w_j(t) = (1/L)\kappa\,[(t - t_j)/h]$, the discrete sums of Eq. (8a) can be replaced by the integrals:

$$B = \frac{1}{2}\int_{-1}^{1} s^2 \kappa(s)\, ds\,, \quad C = \int_{-1}^{1} \kappa(s)^2\, ds\,. \quad (8b)$$

Making a Taylor-series expansion in Eq. (5), we approximate the local bias as

$$\text{Bias}[\hat{g}(t_i)] = \mathbf{E}[\hat{g}(t_i)] - g(t_i) \simeq Bg''(t_i)h^2\,. \quad (9)$$

Equation (9) predicts that the bias increases as $h^2$. The Taylor-series approximation of the bias is reasonably close to the exact bias except when $g''(t)$ nearly vanishes. In this case, higher-order terms need to be included to evaluate the ESE.

Using Eqs. (4) and (8), the local ESE reduces to

$$\text{ESE} \simeq [Bg''(t_i)h^2]^2 + \frac{\sigma^2 C\Delta}{h}\,. \quad (10)$$

If we minimize the local ESE [as given by Eq. (10)] with respect to the kernel halfwidth $h$, the optimal halfwidth is

$$h_{as}(t) = \left[\frac{\sigma^2 C\Delta}{4B^2 |g''(t_j)|^2}\right]^{1/5}\,. \quad (11)$$

For this choice of kernel width $h_{as}$, the total expected squared error of Eq. (10) is

$$\text{ESE} \simeq 1.65\,|Bg''(t)|^{2/5} (\sigma^2 C\Delta)^{4/5}\,. \quad (12)$$

Thus, the optimal $h$ is proportional to $\Delta^{1/5}$, and the total squared error is proportional to $\Delta^{4/5}$. [Equation (12) can be used to give error bars for kernel-smoother estimates.]

Equation (12) is an asymptotic formula and may or may not be a good approximation for a particular signal and data set. In Fig. 5, we compare the local ESE approximation with the actual ESE for the noisy chirp signal. The solid line is the actual value of the ESE for $h = 0.08$, while the dotted line gives the local approximation of the ESE. For $t \le 0.5$, the two curves agree. For larger times, the shapes and magnitudes of the two curves are very similar, but there is a phase shift between them due to higher-order Taylor-series terms.

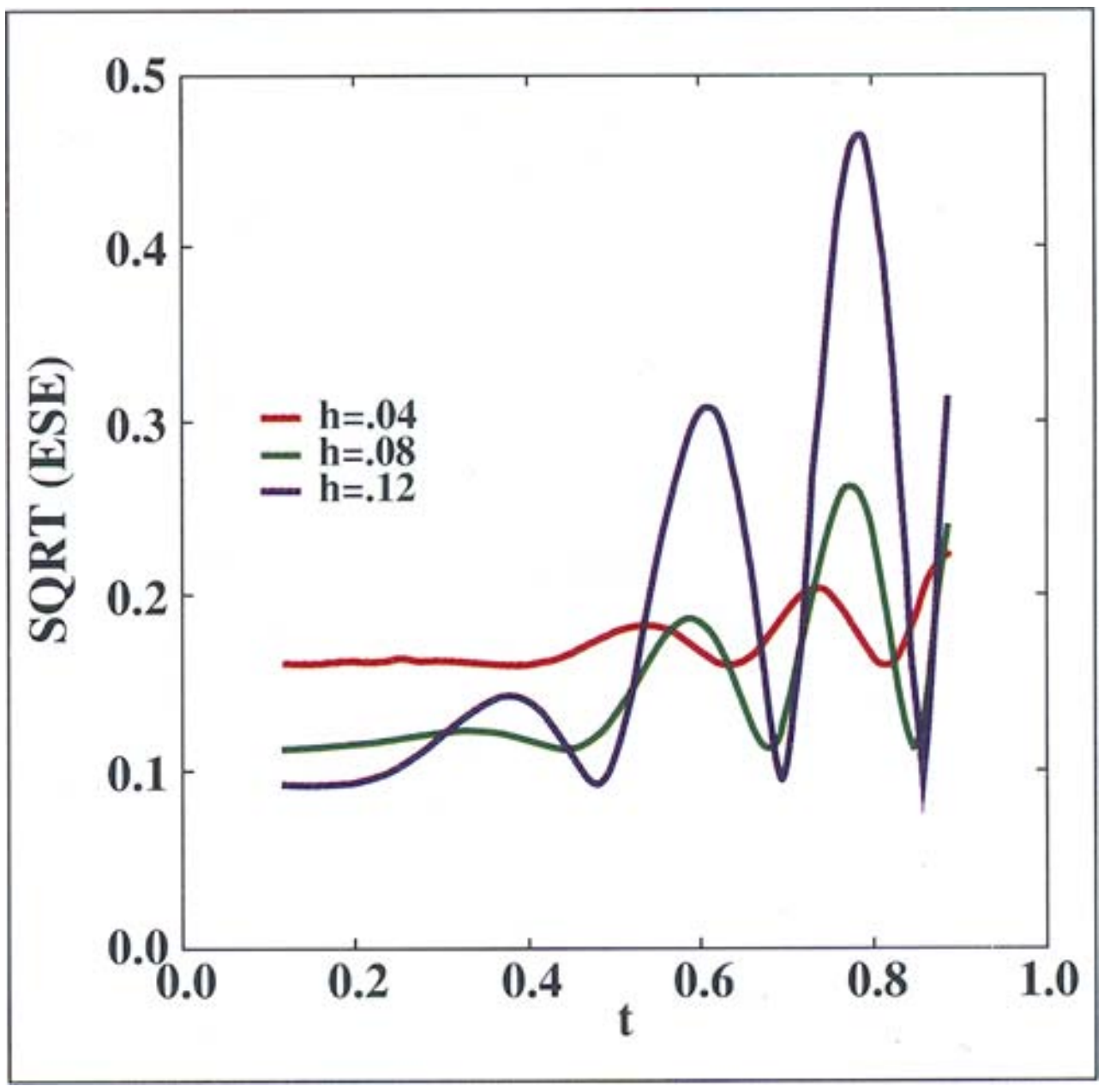

*Figure 4. Expected square error (ESE) of the parabolic-kernel smoothers of Fig. 3. As time progresses, g(t) varies more rapidly, and the bias error grows for all three estimates. As a result, the ESE for all three estimates grows and oscillates proportionally to $|g''(t)|^2$. Each of the halfwidths performs best in a different time interval. The smallest halfwidth has the largest ESE for $t \le 0.5$. For $0.75 \le t \le 1.0$, the smallest halfwidth is the most reasonable.*

Equation (11) drives data-adaptive kernel estimation. In essence, Eq. (11) shows that when $g(t)$ is rapidly varying

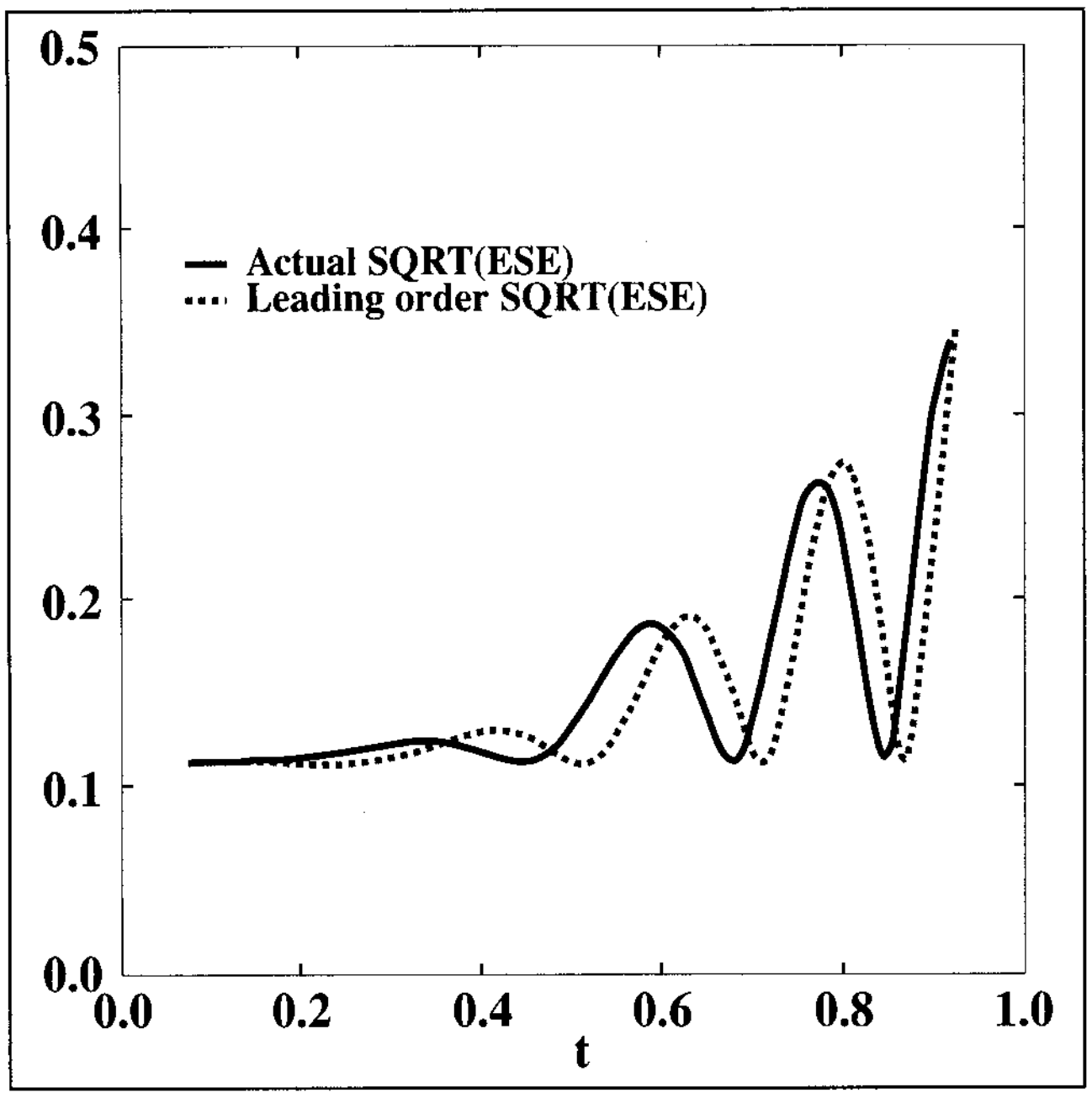

*Figure 5: Comparison of the leading-order ESE given by Eq. (10) with the exact ESE for a kernel-smoother halfwidth of h = 0.08. The shape and magnitude of the two curves are very similar, but there is a phase shift between them due to higher-order Taylor-series terms.*

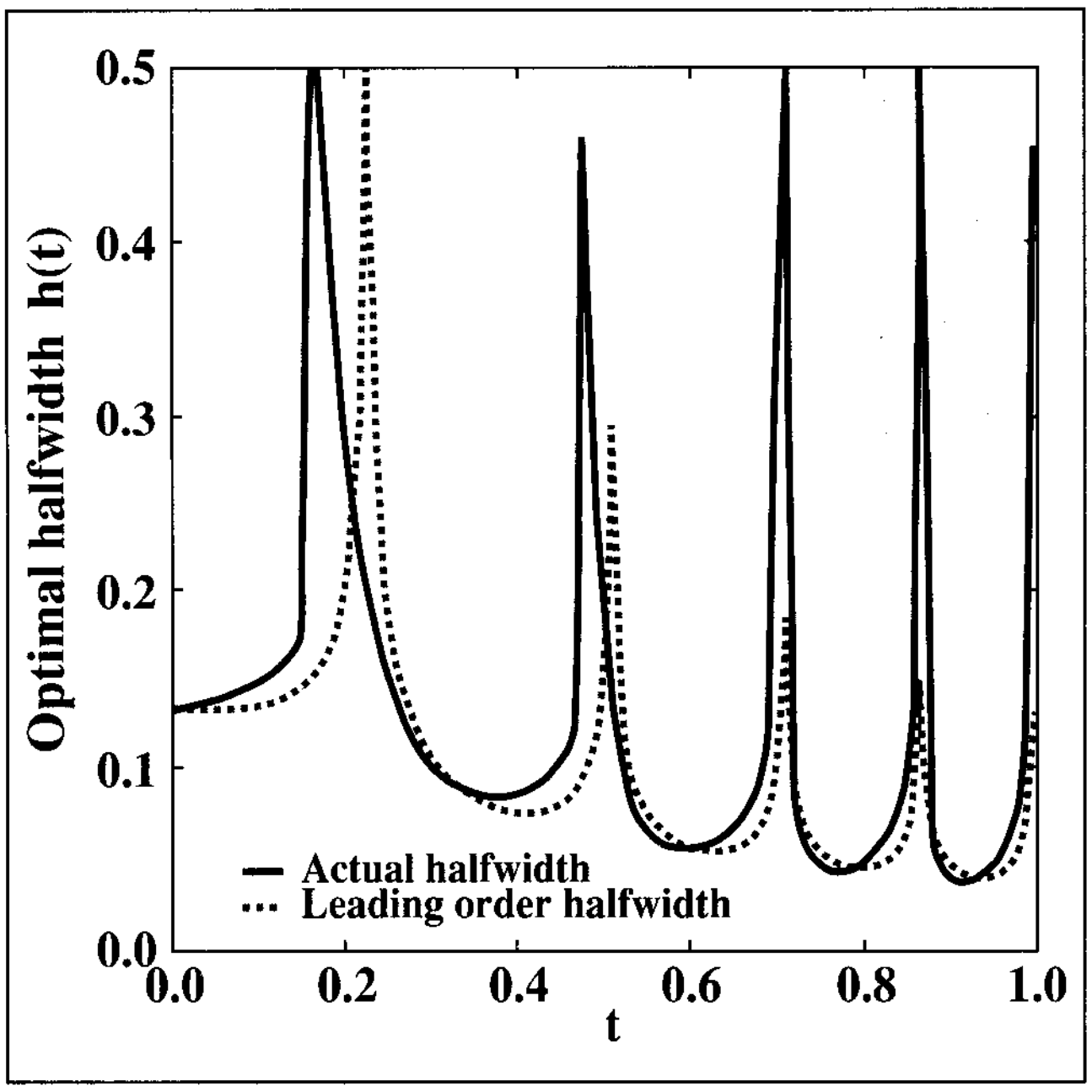

*Figure 6. Optimal halfwidth of kernel smoother as a function of time. The solid line gives the halfwidth that minimizes the exact ESE, while the dashed line gives the halfwidth based on a Taylor-series approximation. At the four extrema of $|g''(t)|$, $h_{as}(t)$ becomes infinite. The phase shift between the extrema of the exact optimal halfwidth and the Taylor-series optimal halfwidth is apparent.*

($|g''(t)|$ is large), the kernel halfwidth should be decreased. Equation (11) gives an explicit solution for the halfwidth that minimizes the local bias-versus-variance trade-off.

Equation (11) has two major difficulties. First, $g''(t)$ is unknown, and thus Eq. (11) cannot be used directly. Estimating $h_{as}(t)$ is considered in the next section. Second, Eqs. (10)-(12) are based on a Taylor-series expansion, and the expansion parameter is $h_{as} \sim \Delta^{1/5}$. Even when $\Delta$ is small, corresponding to fast sampling, $\Delta^{1/5}$ may not be so small.

Figure 6 compares the halfwidth of Eq. (11) with the halfwidth that minimizes the actual ESE. [We calculate the exact ESE by evaluating Eq. (5) for the bias contribution to Eq. (6).] At the four extrema of $|g''(t)|$, $h_{as}(t)$ becomes infinite. Similarly, the actual optimal halfwidth becomes large at four nearby points where the exact bias nearly vanishes. The phase shift between the exact and approximate halfwidths is also apparent.

Figures 5 and 6 illustrate that the local approximation is valuable but can be fallible. Furthermore, the $|g''(t)|^{-2/5}$ dependence of the local optimal halfwidth makes $h_{as}(t)$ depend sensitively on the second derivative of $g(t)$.

## How to select the halfwidth

There are two main approaches to selecting the smoothing parameter $h$ from the data: goodness-of-fit estimators and plug-in derivative estimators. Figure 6 shows that the kernel halfwidth should be adaptively adjusted as the estimation point $t$ varies. In practice, most codes use a fixed halfwidth for simplicity and stability. In this section, we describe two goodness-of-fit estimators. In the next section, we present a plug-in-derivative scheme with variable halfwidth.

*Penalized Goodness-of-Fit Halfwidth Selection.* When the halfwidth is constrained to be constant over the entire interval, we want to choose a constant value of $h$ to minimize the ESE averaged over all sample points, which we denote by EASE for "expected average square error." Since the EASE is unknown, a number of different methods have been developed to estimate it. In the goodness-of-fit methods, the average square residual error (ASR) is evaluated as a function of the kernel halfwidth:

$$\mathrm{ASR}(h) = \frac{1}{N}\sum_i |y_i - \hat{g}(t_i|h)|^2 , \qquad (13)$$

where $\hat{g}(t_i|h)$ is the kernel-smoother estimate using the halfwidth $h$. The ASR systematically underestimates the actual loss, because $y_i$ is used in the estimate $\hat{g}(t_i|h)$ of $y_i$. The variance term in the EASE is

$$V(h) \equiv \sigma^2 \sum_{j=-L}^{L} w_j^2 \approx \frac{\Delta}{h}\sigma^2 C ,$$

where C is given in Eq. (8b). Correspondingly, the variance term in the ASR is

$$\hat{V}(h) = \sum_{j=-L}^{L} (w_j - \delta_{0,j})^2 \approx V(h) + \sigma^2[1 - \frac{2\Delta}{h}\kappa(0)] ,$$

where $\kappa(0)\Delta/h$ is the weight of $y_i$ in the kernel smoother. Thus,

# Special cases: kernel shapes, end points, and unequally spaced data

Kernel smoothers are local averages of the measurement. Thus a key question is "Which weighting is best?" We interpret "best" to mean minimizing the asymptotic ESE as given by Eq. (10). For equispaced data in the continuum limit, the answer is the parabolic weighting: $\kappa(t) = 3/4(1 - t^2)$. This parabolic weighting is optimal provided that the kernel halfwidth satisfies Eq. (11). More generally, to estimate the $q$th derivative to $O(h^2)$, the limiting shape of the optimal kernel is

$$\kappa(t) = \frac{(2q+1)!}{2^{q+1}(q!)}[P_q(t) - P_{q+2}(t)], \qquad \text{(A1)}$$

where $P_q(t)$ and $P_{q+2}(t)$ are the Legendre polynomials. To estimate the second derivative, Eq. (A1) reduces to $(105/16)(-1 + 6t^2 - 5t^4)$. In (A1), the estimation point is at $t = 0$ and the kernel support is [-1,1].

### *Does the kernel shape really matter?*

For the parabolic kernel, the kernel parameters in Eq. (8b) are $B = 1/10$ and $C = 3/5$. In contrast, the rectangular kernel has $B = 1/6$ and $C = 1/2$. From Eqs. (11)-(12), the optimal halfwidth for the parabolic kernel is 27% longer than that of the rectangular kernel, but the ESE of the rectangular kernel is 6% larger than that of the parabolic kernel. The parabolic kernel outperforms the rectangular kernel because the bias error increases rapidly as the $t_{i+j} - t_i$ increases. The parabolic kernel compensates for this increasing bias by downweighting the points that are farthest away from the estimation point. More importantly, a rectangular window causes the estimate $\hat{g}(t)$ to be discontinuous in $t$ while the parabolic kernel estimate is continuous in $t$.

### *What if the data points are not equispaced?*

When the times between data points $t_{i+j} - t_i$ vary, the kernel shape needs to be modified to enforce the moment conditions. A simple way to do this is to reformulate the kernel smoother as a local linear (polynomial) regression. In other words, we fit the data $\{y_{i-L} \ldots y_{i+L}\}$ to a local polynomial, in this case $y_{i+j} \sim g(t) + g'(t)(t_{i+j} - t)$. We need to estimate both $g(t)$ and $g'(t)$ from the data. We do this by a weighted local regression:

$$\{\hat{g}(t), \hat{g}'(t)\} = \text{argmin}_{\{a,\, b\}} \sum_{j=-L}^{L} W_j \,|a + b\,(t_{i+j} - t) - y_{i+j}|^2, \qquad \text{(A2)}$$

where $W_j$ are arbitrary weighting factors. When the $W_j$ are given a parabolic weighting, the local-polynomial-regression estimate is identical to the kernel-smoother estimate with a parabolic weighting in the equispaced case.

The local-polynomial-regression formulation has three advantages. First, it automatically enforces the moment conditions even when the data are not equispaced. Second, it can be quickly updated as the estimation point is advanced in time. Third, polynomial regression works well in two- and three-dimensional settings because it automatically handles boundary effects.

### *How should the kernel weighting be modified near the boundary?*

The parabolic kernel is optimal in the center of the domain. When the estimation point $t_i$ is at or near the boundary, a nonsymmetric kernel should be used. We define the touch point $t_p$ as the point where $h_{as}(t_p) = t_p$ [with $h_{as}$ given in Eq. (11)]. When the estimation point is beyond the touch point, we fix the the kernel halfwidth to $h = h_{as}(t_p)$. We then adjust the kernel shape to maintain the moment conditions while keeping the ESE small.

We consider kernel estimates of the $q$th derivative near the left boundary of the data. We normalize the kernel weights near the boundary:

$$w_j(t) = \frac{\Delta}{h^{q+1}} G\left(\frac{t}{h} - 1,\ \frac{t_j}{h} - 1\right).$$

The function $G(z, y)$ is defined for $-1 \le y \le 1$ and $-1 \le z \le 0$. In the continuum limit, the boundary kernel that minimizes the ESE [under the assumption that $h(t) = h_0(t)$, i.e., that the kernel halfwidth is optimal] is

$$G(z, y) = \frac{(2q+1)!}{2^{q+1}(q!)}\{P_q(y) + (2q+3)\,z\,P_{q+1}(y) + [(2q+3)]\,z^2 - 1)\,P_{q+2}(y)\}. \qquad \text{(A3)}$$

When the data points are not equispaced, the simplest approach is to use local polynomial regression with the halfwidth fixed at $h = h_{as}(t_p)$. A simple weighting near the boundary is a parabola centered about the estimation point:

$$W_j(t) = 1 - \left(\frac{t - t_j}{t - 2t_p}\right)^2.$$

### *What if the variance $\sigma^2$ is unknown?*

We estimate it through three point residuals:

$$\hat{\sigma}^2 = \frac{2}{3(N-2)} \sum_{i=2}^{N-1} \left(y_i - \frac{y_{i+1} + y_{i-1}}{2}\right)^2. \qquad \text{(A4)}$$

Equation (A4) neglects the bias error from $g''(t)$. This is reasonable because only adjacent points are used.

$$\mathbf{E}[\mathrm{ASR}(h)] = \mathrm{EASE} + \sigma^2\left(1 - \frac{2\kappa(0)\Delta}{h}\right). \qquad (14a)$$

Goodness-of-fit methods determine the kernel halfwidth

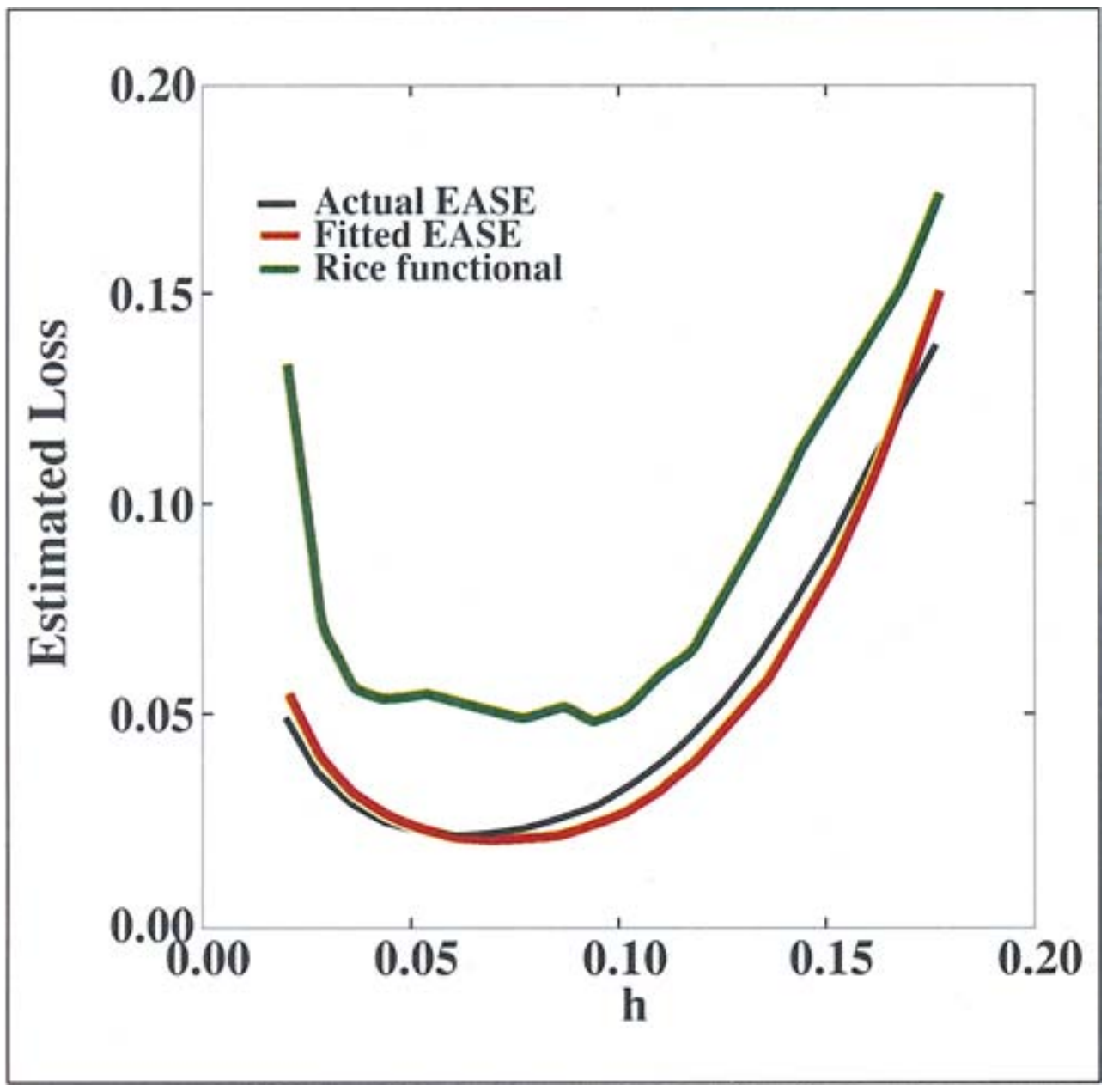

*Figure 7. Empirical fits to the expected average square error. The solid red curve is the actual EASE as a function of kernel halfwidth. The solid black curve is the Rice functional minus $\sigma^2$. The dotted green curve is the estimated EASE using the two-parameter fit to the ASR. In this plot, only the locations of the minima are important and not the quality of the fit. Random oscillations can shift the location of the minimum of the corrected ASR appreciably. The minimum of the fitted ASR criterion is less sensitive.*

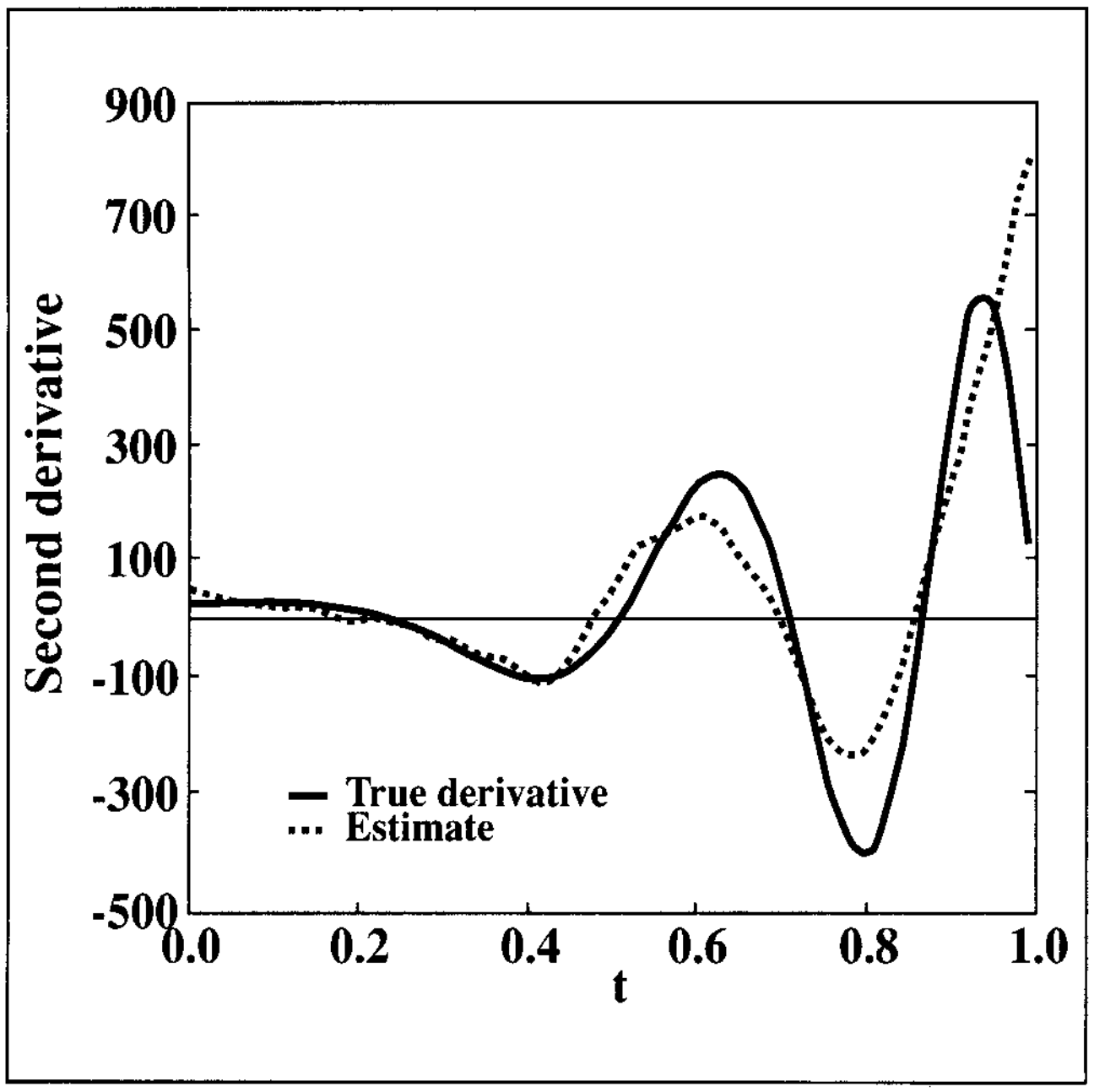

*Figure 8. Comparison of g″ (t) (red line) with its estimate (black line). Since the halfwidth is fixed at h = 0.15, we undersmooth the initial times and oversmooth the more rapidly varying segment near the end of the data.*

$h$ by minimizing an empirical estimate of the EASE. A number of different goodness-of-fit functionals have been proposed. A simple goodness-of-fit functional is given by "inverting" Eq. (14a) under the ansatz that $\mathrm{ASR}(h) \simeq \mathbf{E}[\mathrm{ASR}(h)]$ and that $\mathrm{ASR}(h) \simeq \sigma^2$. The resulting estimate of the EASE is

$$\widehat{\mathrm{EASE}} = \mathrm{ASR}(h)\left(1 + \frac{2\kappa(0)\Delta}{h}\right). \qquad (14b)$$

Monte Carlo tests have shown that minimizing Eq. (14b) with respect to $h$ tends to underestimate the value of the optimal halfwidth. As a result, a number of alternative estimates have been proposed. We consider one of the more popular alternatives, the Rice criterion, which selects the kernel halfwidth $h_{Rice}$ by minimizing

$$C_R(h) = \mathrm{ASR}(h)\left(1 - \frac{2\kappa(0)\Delta}{h}\right)^{-1}. \qquad (14c)$$

Unfortunately, $C_R(h)$, like most of the goodness-of-fit functionals, is often very flat, and its actual minimum can be very sensitive to noise (see Fig. 7). As a result, the halfwidth given by the Rice criterion tends to vary appreciably even when the noise is weak.

*Fitted Mean-Square-Error Halfwidths.* We now consider methods that produce less-sensitive estimates of the best halfwidth. To reduce the random errors in the $h$ optimization, we fit the $\mathrm{ASR}(h)$ with the parametric model:

$$\mathrm{ASR}\,(h) \sim a\hat{V}(h) + bh^4. \qquad (15)$$

The model has two free parameters, $a$ and $b$. The first one corresponds to $\sigma^2$, and the second one corresponds to $|Bg''(t)|^2$. By parameterizing $\mathrm{ASR}(h)$, we are assured of a unique minimum. We determine $a$ and $b$ by minimizing the least-squares problem:

$$\{a,b\} = \mathrm{argmin}_{\{a,b\}} \sum_{h_j} \{\mathrm{ASR}\,(h_j) - [a\hat{V}(h_j) + bh_j^4]\}^2, \quad (16)$$

where the $h_j$ are equispaced in $h$ with a cutoff value of $h_j$ chosen such that $C_R(h_{cutoff}) \sim 2C_R(h_{Rice})$. We then select the halfwidth to minimize $aV(h) + bh^4$. This fitting procedure gives more stable halfwidth estimates than the penalized goodness-of-fit methods do.

The solid red line in Fig. 7 is the actual EASE. The black line is the Rice functional minus $\sigma^2$. The dashed green line is our estimate of the EASE based on the two-parameter fit to the ASR. Near the minimum, random oscillations can shift the location of the minimum by an appreciable amount. This effect is even more pronounced when the correction term is estimated empirically. The minimum of the fitted ASR is closer to the actual optimal halfwidth due to the stabilizing effect of the $h$ values with larger EASE. Note that in this plot only the locations of the minima are important and not the quality of the fit.

## Plug-in-derivative estimates of the local halfwidth

In this method, we begin by estimating $g''(t)$ and then inserting this estimate into Eq. (11). This estimation and insertion procedure is called a plug-in-derivative estimate. We describe a particular plug-in method based on the work of Müller and Stadtmüller. The scheme is somewhat complicated because it is optimized to give optimal performance when the sampling rate is very high.

To estimate $g''(t)$, we use a kernel that satisfies

$$\frac{1}{2}\sum_{j=-L}^{L}(t_{i+j}-t_i)^m\,w_j=\begin{cases}0 & \text{if } m=0,1,3;\\ 1 & \text{if } m=2.\end{cases}$$

Equation (A1) presents an "optimal" kernel for estimating derivative. The bias in the estimating $g''(t)h^2$ scales as $g^{(iv)}h^4$, while the variance decreases as $\Delta/h$. Optimizing the bias-versus-variance trade-off yields an optimal halfwidth for estimating second derivatives $h_2$, which is proportional to $\Delta^{1/9}$. The halfwidth for estimating $g''(t)$ is larger than that of the halfwidth for estimating $g(t)$ because the bias error increases more slowly ($h^4$ versus $h^2$).

Now we may plug the estimates of $g''(t)$ and $\sigma^2$ into Eq. (11) to get the variable halfwidth $h_{as}(t)$ for estimating $g(t)$. The described procedure is not complete because the halfwidth for estimating $g''(t)$ is not specified. Since the final estimate is not sensitive to the initial halfwidth, it is possible to input the halfwidth by hand. A better approach is to begin the estimation procedure using a goodness-of-fit estimator. In our numerical implementation, we use the fitted goodness-of-fit method to determine the halfwidth for estimating $g''(t)$. When the goodness-of-fit initialization is included, the combined method has three stages:

1. Goodness-of-fit initialization to determine $h_2$.
2. Kernel-smoother estimation of $g''(t)$.
3. Final kernel estimation using the plug-in estimate of $h_{as}(t)$.

To illustrate this multiple-stage estimation, Fig. 8 plots the second-derivative estimate of the chirp signal. The fitted ASR criterion yielded an optimal halfwidth of 0.15 for the estimate of $g''(t)$. Since this halfwidth is time-independent, we undersmooth the initial times and oversmooth the more rapidly varying segment near the end of the data.

Figure 9 gives the plug-in estimate of the optimal halfwidth for smoothing $g(t)$. When $g''(t)$ is zero, the "optimal" halfwidth $h_{as}(t)$ is infinite because Eqs. (10)-(12) neglect the higher-order bias, which in this case is proportional to $g^{(iv)}(t)h^4/(4!)$. We regularize Eq. (12) by replacing

$$|\widehat{g''}(t)|^2$$

with

$$|\widehat{g''}|^2 := |\widehat{g''}(t)|^2 + k_1\left|\frac{\widehat{g^{(iv)}}(t)h^2}{12}\right|^2, \qquad (17)$$

where $k_1$ is a constant. To estimate the magnitude of

$$|\widehat{g^{(iv)}}(t)|^2,$$

we invert the analog of Eq. (11) for the second-derivative estimate:

$$|\widehat{g^{(iv)}}|^2 \sim |\widehat{h_2}|^9 .$$

Similarly, we eliminate $h^2$ from Eq. (17) using Eq. (11). The resulting expression is

$$|\widehat{g''}|^2 \leftarrow |\widehat{g''}(t)|^2 + k_2\,|\widehat{h_2}|^5 .$$

We may use $k_2$ as a tuning parameter for the regularization of the kernel halfwidth. The red line in Fig. 9 shows the regularized halfwidth.

Figure 10 displays the final estimate of the chirp signal. The dotted lines are the error bars for the final estimate [obtained by substituting our smoothed estimate of $|g''(t)|^2$ into Eq. (12)].

Figure 11 continues the ESE comparison of Fig. 4. The red curve is the ESE of the multiple-stage variable-halfwidth kernel estimator. The other three curves are the fixed-halfwidth parabolic-kernel ESEs. The plug-in halfwidth automatically decreases the halfwidth where $|g''(t)|$ is large. As a result, the ESE of the multiple-stage estimate is comparable to the $h = 0.12$ fixed-halfwidth kernel for early times and to the $h = 0.04$ fixed-halfwidth kernel for later times. Essentially, the data-adaptive multistage estimate gives an ESE that is comparable to the best possible fixed-halfwidth ESE locally.

## Data-adaptive smoothing

Local averaging is a common data-analysis technique. However, most of the practitioners make arbitrary and suboptimal choices about the amount of smoothing. In this tutorial, we determined the expected error by smoothing the data locally. Then we optimized the shape of the kernel smoother to minimize the error. Because the optimal estimator depends on the unknown function, our scheme automatically adjusts to the unknown function. By self-consistently adjusting the kernel smoother, the total estimator adapts to the data.

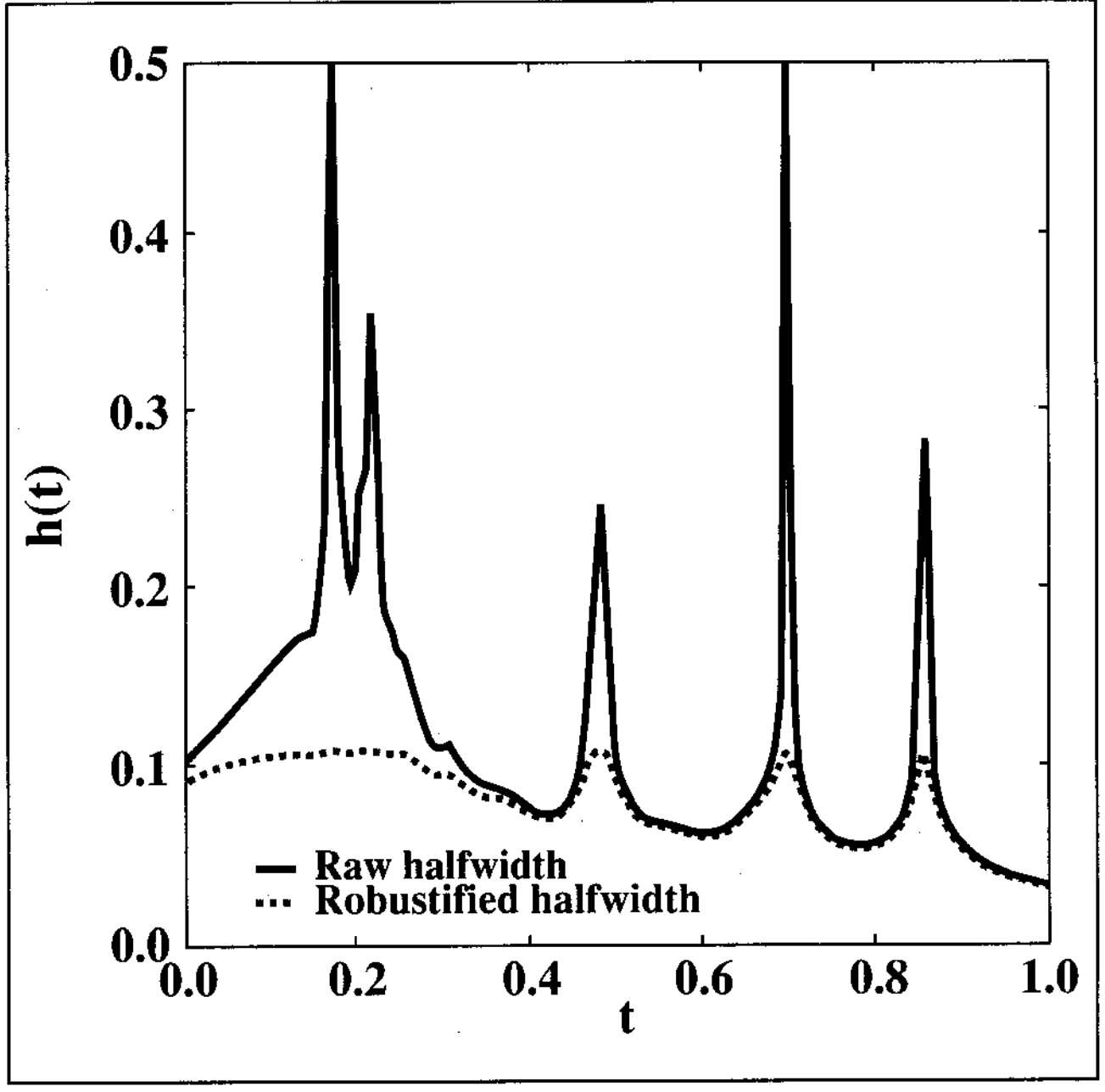

*Figure 9. Inferred value of the optimal kernel halfwidth using the plug-in estimate of Eq. (11). The solid black line is the simple plug-in, and the dotted red curve is the regularized plug-in using Eq. (17).*

Goodness-of-fit estimators select a kernel halfwidth by minimizing a function of the halfwidth that is based on the average square residual fit error, ASR($h$). A penalty term is included to adjust for using the same data to estimate the function and to evaluate the mean-square error. Goodness-of-fit estimators are relatively simple to implement, but the minimum (of the goodness-of-fit functional) tends to be sensitive to small perturbations. To remedy this sensitivity problem, we fit the mean-square error to a two-parameter model prior to determining the optimal halfwidth.

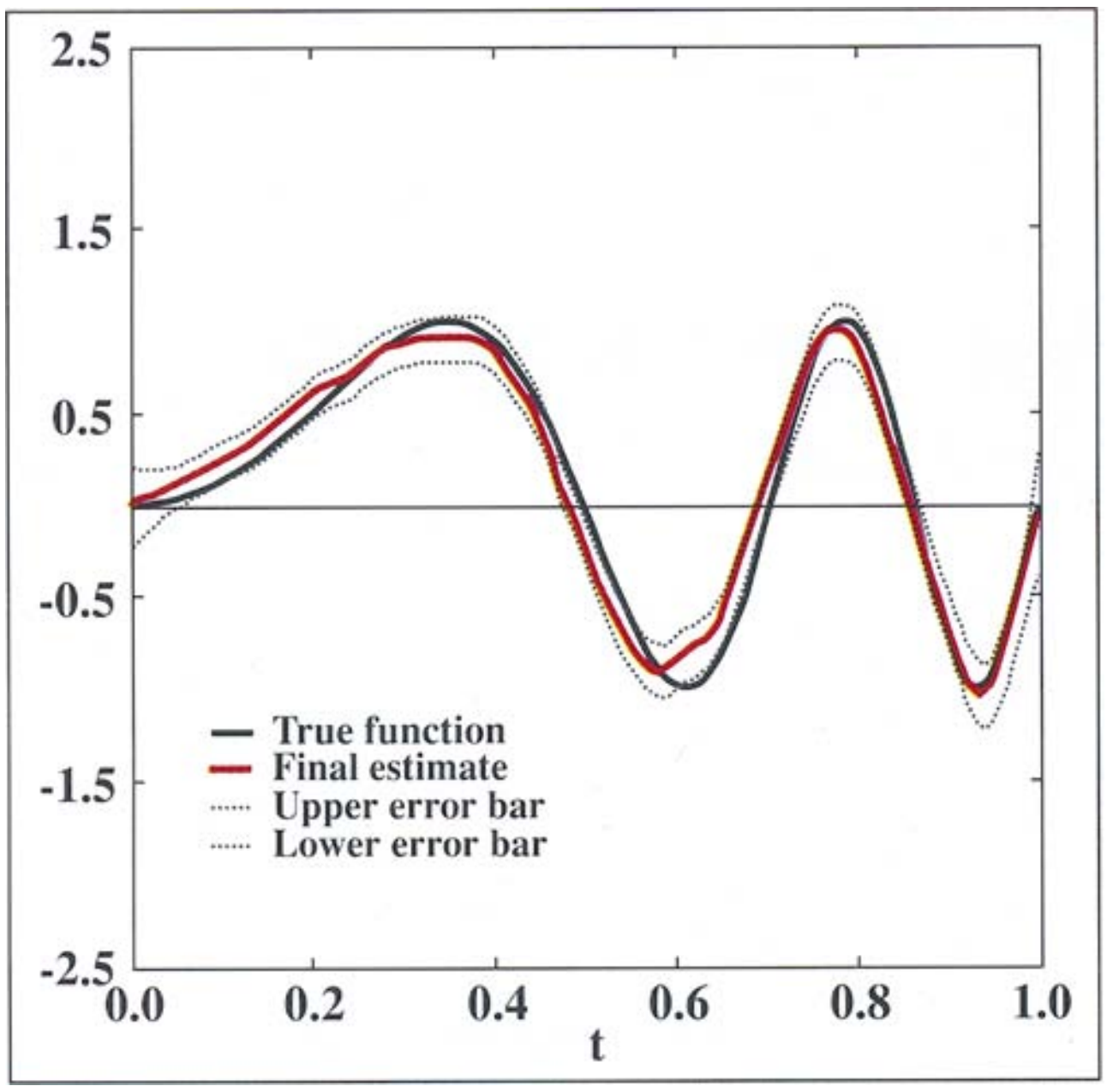

*Figure 10. Final estimate of the chirp signal using the multistep estimation procedure. The error bars are evaluated using Eq. (12).*

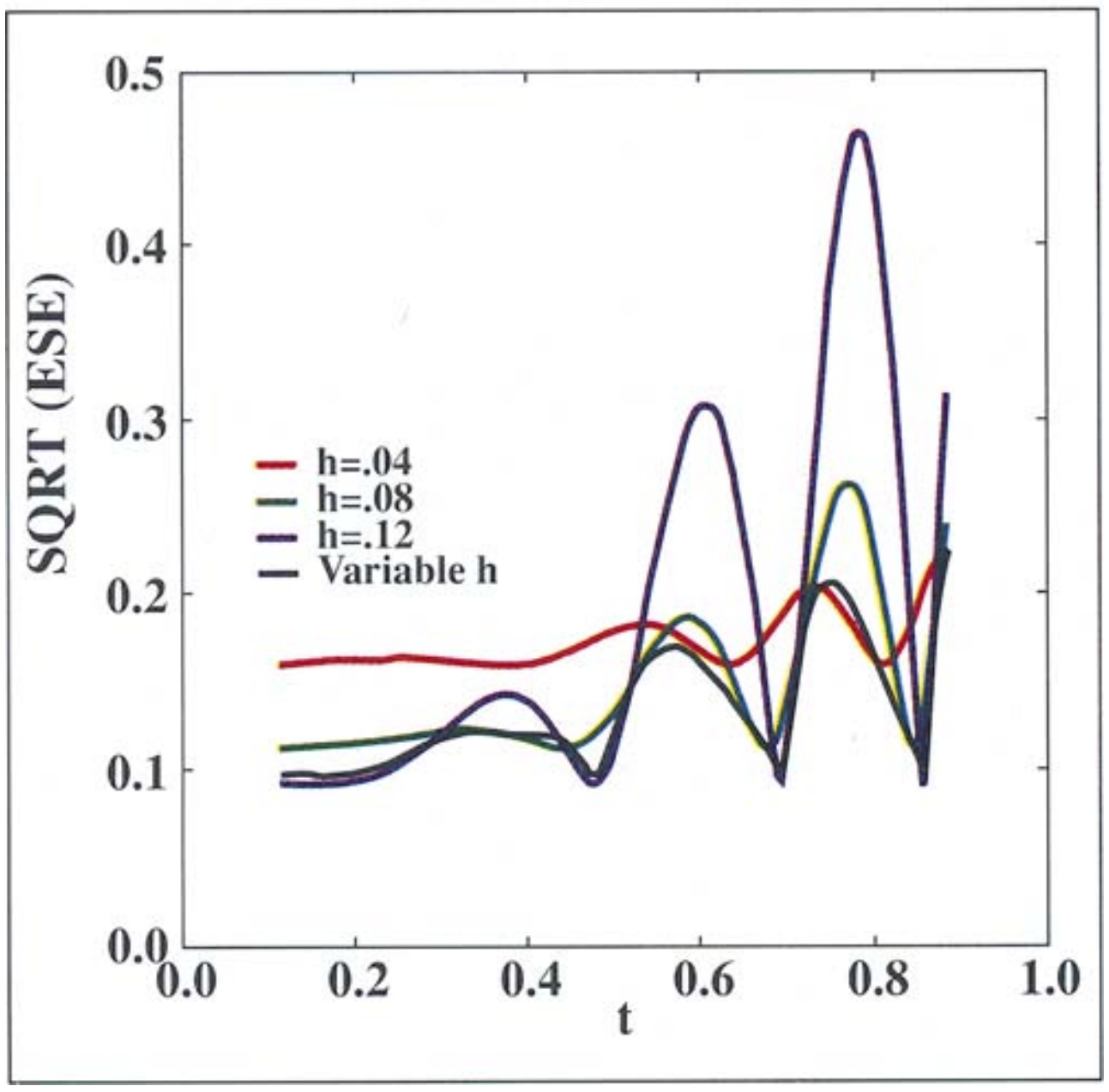

*Figure 11. Comparison of the ESE of the data-adaptive variable-halfwidth kernel smoother with the fixed-halfwidth kernel smoothers. The red curve is the ESE of the multiple-stage kernel estimator. The other three curves are the fixed-halfwidth parabolic-kernel ESEs. The plug-in halfwidth automatically decreases the halfwidth where $|g''(t)|$ is large. As a result, the multiple-stage-estimate's ESE is comparable to the smallest of the three ESEs of the fixed-halfwidth kernels at all time points.*

Plug-in-derivative estimators approximate the second derivative of the unknown function in an initial step, and then substitute this estimate into the asymptotic formula in Eq. (11). The multistage plug-in method is optimized to give the best possible performance when the sampling rate is very high. Figure 9 illustrates that the quality of fit is visibly enhanced using this plug-in estimate.

We caution that the analysis is based on a Taylor-series approximation of the unknown function. Essentially, we require that the sampling rate is sufficiently rapid in comparison with the time scale on which the unknown function is varying.

In the box, we have discussed two practical problems: estimation near the ends of the data and unequally spaced data. If your data are equispaced, the parabolic kernel in the interior should be modified to the edge kernel [see Eq. (A3)] near the ends of the data. If your data are not equispaced use local linear regression [see Eq. (A2)]. To reduce computation cost, the regression matrix can be modified by rank-one updates as the estimation point is advanced.

## Further reading

Two textbooks on kernel estimation are Hardle (1990)[1] and Müller (1980).[2] The local-polynomial-regression approach is described in Hastie and Loader (1993).[3] The equivalence of kernel smoothers and kernel regression is given in Sidorenko and Riedel (1993).[6] This article also describes kernel estimation near the ends of the data. The goodness-of-fit approach to selecting kernel halfwidths is discussed in Hardle, Hall, and Marron (1988).[4] The multistage data-adaptive scheme is described in Müller and Stadtmüller (1987)[5] and in Brockman, Gasser, and Hermann (1994).[7]

## Acknowledgment

The presentation has benefitted from a critical reading by Joe Keller. We thank Jim Matey for reformatting the figures. This work was funded by the U.S. Department of Energy, Grant DE-FG02-86ER53223.